\documentclass[a4paper,11pt]{article}
\usepackage{pos}
\usepackage{needspace}
\usepackage{tabularx}
\usepackage[skip=1ex]{caption}
\usepackage{wrapfig}
\title{Ukrainian contribution to particle physics: historical perspective and prospects}

\author*[a]{Denys Timoshyn}
\author[c]{Tetiana Hryn'ova}
\author[b]{Igor Kostiuk}

\affiliation[a]{Institute of Particle and Nuclear Physics, Faculty of Mathematics and Physics, Charles University, V Holešovičkách 2, 180 00 Prague 8, Czech Republic}

\affiliation[b]{NIKHEF, Science Park 105, 1098 XG
Amsterdam, NH, Netherlands}

\affiliation[c]{LAPP, Université Savoie Mont Blanc, CNRS/IN2P3, 9 Chemin de Bellevue, Annecy,  France}

\emailAdd{denys.timoshyn@cern.ch}
\emailAdd{tetiana.hryn'ova@cern.ch}
\emailAdd{igor.kostiuk@cern.ch}

\abstract{Many world-known scientists and engineers like G. Breit, G. Budker, G. Charpak, G. Gamow, M. Goldhaber, A. Ioffe, S. Korolyov, E. Lifshitz, M. Ostrogradsky, S. Timoshenko, V. Veksler were born in Ukraine, while some, like L. Landau and M. Bogolyubov, started their career there. Reclaiming their scientific legacy as well as that of many others helps to promote Ukrainian contributions to particle physics both inside and outside of Ukraine and to motivate the next generation of Ukrainian scientists in the time of war. We will present the status of Ukrainian scientific infrastructure two years after the start of the full-scale invasion and past, present and expected future contributions of Ukrainian scientists to CERN.}

\FullConference{42nd International Conference on High Energy Physics (ICHEP2024)\\
18-24 July 2024\\
Prague, Czech Republic\\}

\begin{document}
\maketitle

\section{Historical Overview}

Scientists from Ukraine contributed to the field of particle physics from its conception, yet their contributions are rarely associated with Ukraine. One potential reason: many foreign scientists were oblivious to the existence of Ukraine and Ukrainians prior to the modern declaration of independence in 1991, despite the creation of the Ukrainian Academy of Sciences in 1918 and the Ukrainian SSR status as one of the founding members of the United Nations (1945).

\begin{wrapfigure}[20]{r}{0.6\textwidth}
\centering
\includegraphics[width=\linewidth]{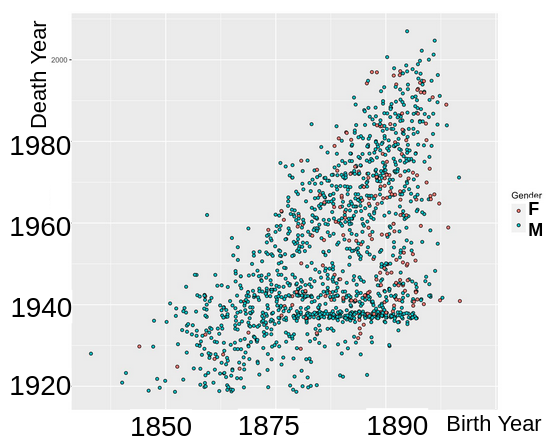}  
\caption{Graph of birth and death years of Ukrainian scientists associated with the Ukrainian Academy of Sciences. 
The pattern at the year 1937 coincides with The Great Purge done by the Soviet regime. 
\href{https://www.facebook.com/olexii.ignatenko/posts/pfbid02dg1s4yPsC29epmFYdgZhZhXFtFx57twKNwkJe9HGgkgD6dns5iNbQcsadCQ6T4fol}{(From O. Ignatenko)}
}
\label{birthDeathYears}
    \end{wrapfigure}
In 1972 V. Weisskopf wrote~\cite{vw} about his work at the \textbf{\textbf{Ukrainian} Institute of Physics and Technology (UPTI)} in Kharkiv, then capital of the Ukrainian Soviet Socialist Republic, as follows: ``\textit{In 1933 I went to Kharkov, \textbf{Russia} for almost a year, where it was possible to get a job with subsistence. Working in Kharkov at that time were Landau, Lifschitz, and Achiezer, and many other young Russian physicists.}'' It is disappointing to see Landau, Lifschitz, and Akhiezer, labeled as Russian physicists. Landau was born in Baku, his parents coming from present-day Belarus. Lifschitz was born in Kharkiv. Akhiezer was born in present-day Belarus, but worked all his life in Kharkiv. Indeed, UPTI, in Kharkiv, was the place to be at the time: Landau and Lifschitz started to write their ten-volume course on Theoretical Physics there, Bohr, Van der Graaff, Dirac, among many other foreign scientists, visited for conferences or work~\cite{kipt1}. 
UPTI was founded in 1928 at the initiative of ``the father of Soviet physics'' Abram Ioffe (born in Romny, now in Ukraine), who launched institutes of Physics and Technology in many industry centers of the USSR. The Soviet government was wary of research freedom and international collaborations of Ukrainian scientists. In the late 1930s, 16 scientists from UPTI were arrested, including Landau, 8 of them were subsequently executed, like many other Ukrainian scientists at the time (see Figure~\ref{birthDeathYears}), and all foreign scientists were forced to leave the Ukrainian SSR. The UPTI, renamed \href{https://www.kipt.kharkov.ua/en.html}{Kharkiv Institute of Physics and Technology (KIPT)}, after the World War II was chosen to host Soviet Laboratory No. 1 for nuclear weapons development~\cite{kipt1}, putting a layer of secrecy on the work done there. Nevertheless, it continued to play a leading role as a Ukrainian center for particle physics research with notable results being the work of 
\begin{itemize}
   \item Akhiezer and \href{http://archive.ujp.bitp.kiev.ua/files/journals/52/10/521014p.pdf}{Shulga} on the development of quasi-classical theory of coherent radiation of channelled and over-barrier electrons and positrons in crystals. Ternovsky-Shulga–Fomin effect and Grinenko–Shulga mechanism have recently been confirmed experimentally by NA63 and UA9 experiments at CERN.
    \item Volkov, Akulov and Soroka from KIPT played \href{https://cerncourier.com/a/the-many-lives-of-supergravity/}{a crucial role} in the development of supergravity and supersymmetry. These two theories are eagerly pursued for an experimental validation. 
\end{itemize}
Kharkiv also plays important role in the development of scintillating materials for particle physics. In 1992, \href{https://en.wikipedia.org/wiki/Ludmyla_Nagorna}{Ludmyla Nagorna} 
(\href{https://www.scopus.com/authid/detail.uri?authorId=56231721900}{also known as Ludmila Nagornaya}), from the Institute for Single Crystals in Kharkiv, proposed to use lead tungstate crystals (PbWO$_4$) for particle physics experiments. They are now at the heart of calorimeters of the ALICE and CMS detectors at the LHC, CERN. Kharkiv scintillators are also collecting data for the electromagnetic calorimeter of the Belle II detector at KEK (Japan), Fermi LAT space telescope, and hadronic calorimeter of the CMS experiment at CERN. They were used in the electromagnetic calorimeter of the BaBar experiment at SLAC (USA) and in the Minimum Bias Trigger System of the ATLAS detector at CERN. 

Recently, Kharkiv \href{http://isma.kharkov.ua/en}{Institute for Scintillation Materials (ISMA)} in collaboration with KIPT and Kyiv University (TSNUK) have contributed to technology proposal, simulation, prototyping, beam tests and innovative aging studies for the new luminometer (PLUME) for the LHCb detector at CERN. ISMA is also currently coordinating \href{https://www.twisma.eu/}{Horizon Europe TWISMA project} with the participation of CERN and Institute of Light and Matter, University Claude Bernard Lyon, France, focused on research on innovative calorimeters for high energy physics based upon advanced scintillation materials.

Another key center of particle physics in Ukraine is the Bogolyubov Institute for Theoretical Physics, which was established in Kyiv in 1966 by \href{https://ujp.bitp.kiev.ua/index.php/ujp/article/view/2023473}{Mykola Bogolyubov}. It was Bogolyubov himself who brought the ICHEP to Kyiv in 1959 and 1970. This institute has been at the forefront of research in the areas of quantum field theory, statistical mechanics, and nonlinear dynamics. Its scientists have made significant contributions to the ALICE detector at CERN.

The above developments are even more remarkable considering that they were achieved in the climate of persecution and mass emigration of Ukrainian intellectuals. At the same time, the emigration shielded many scientist born in Ukraine from wars, Soviet repressions and famines, allowing them to develop to the full potential for the benefit of the field. Below are a few of our notable compatriots, whose names are rarely associated with Ukraine, with their main achievements. 

 Maurice Goldhaber, born in 1911 in the city of Lemberg (now Lviv). The first of his notable experiments is the photodisintegration of the deuteron (1934) which allowed a measurement of the mass of the neutron. With his wife Gertrude Scharff-Goldhaber, he established in the 1940s that beta particles are identical to electrons. Another major discovery was the determination of the helicity of the neutrino (1958). He was a patriarch of a family that spans three generations of the USA physicists.


 Gregory Breit, born in 1899 in Mykolaiv, over the course of his career in the USA, made pioneering contributions to the theoretical understanding of the nuclear structure and particle dynamics in addition to his highly significant work in atomic and ionospheric physics. The relativistic Breit–Wigner distribution, the Dirac–Coulomb–Breit equation, the Breit frame of reference, the Breit–Wheeler process (which was a subject of  \href{https://cds.cern.ch/record/2895091/files/HIN-21-015-pas.pdf}{recent studies} by the CMS collaboration at CERN) all carry his name.

 George Gamow, born in Odesa in 1904, managed to flee from the Soviet Union to the USA in 1933. He is ``renowned for developing the `Big Bang Theory' of the universe (1948), explaining nuclear alpha decay by quantum tunneling (1928); introducing the `Gamow' factor in stellar reaction rates and element formation (1938); modeling red giants, supernovae, and neutron stars (1939); first suggesting how the genetic code might be transcribed (1954); and popularizing science through a long series of books, including the adventures of `Mr. Tompkins' (1939-1967)''~\cite{gamow}. Gamow's mother was a niece of \href{https://uk.wikipedia.org/wiki/%D0%9B%D0%B5%D0%B1%D0%B5%D0%B4%D0%B8%D0%BD%D1%86%D0%B5%D0%B2_%D0%A4%D0%B5%D0%BE%D1%84%D0%B0%D0%BD_%D0%93%D0%B0%D0%B2%D1%80%D0%B8%D0%BB%D0%BE%D0%B2%D0%B8%D1%87}{F. Lebedintsev}, founder and editor of one of the first Ukrainian monthly historical, ethnographic and literary magazines, \href{https://en.wikipedia.org/wiki/Kievskaia_starina}{Kyivska Staryna (Kyiv's past)}.

 Dmitri Ivanenko, born in 1904 in Poltava, was the first director of the UPTI theoretical division (1929-30). His achievements include: prediction of synchrotron radiation (in collaboration with I.~Pomeranchuk), the proton–neutron model of the atomic nuclei (1932), the first shell model of the nuclei (1932, in collaboration with E. Gapon), the first model of nuclear exchange forces (1934, in collaboration with I. Tamm). His later work focused on the theory of hypernucleus and gravitational theory. His sister, \href{https://en.wikipedia.org/wiki/Oksana_Ivanenko}{Oksana Ivanenko}, was a Ukrainian children's writer and translator.

Gersh Budker was born in 1918 in Murafa, near Vinnytsia in the Ukrainian State. He worked on nuclear, plasma and accelerator physics and is best-known for inventing a method to use electrons to cool heavier particles, which was a major step towards construction of proton-antiproton colliders, and proposing experiments on colliding beams. He founded the Budker Institute of Nuclear Physics in Novosibirsk. 

Vladimir Veksler, born in 1907, Zhytomyr, is most known for his work on accelerators. He, independently of Edwin McMillan, proposed to bunch particles by accelerating them on the edge of the electric field rather than the peak. In 1956 he established and became the first director of the Laboratory of High Energies at the Joint Institute for Nuclear Research. 

Last but not least, Georges Charpak, born in 1924 in Sarny in Polish Republic (now in Ukraine), invented multi-wire proportional chamber, launching the electronic era of particle physics detectors. He was awarded a Nobel Prize in Physics in 1992 for this invention.  

\section{Current challenges: Scientific life during the war}

\subsection{Impact of the war in Ukraine}

Since 2014 Ukraine has been under attack by the Russian Federation. During this time the situation for the science in Ukraine has been challenging. All institutions in the occupied territories were forced to move to other areas of Ukraine, losing all their material base and part of their personnel. Many research institutions, particularly those located in the areas of intense fighting, have suffered severe damage: 1443 buildings belonging to 177 scientific institutions were damaged together with 643 pieces of research equipment \cite{UNESCO}. Around 12\% of Ukrainian scientists (10429) from 524 universities or research institutes were either relocated internally (6.3\%) or moved abroad (5.5\%). Nearly 30\% of Ukrainian scientists were forced to work remotely. At least 1518 scientists volunteered for the military service and more than 70 died either defending Ukraine or as civilian victims of the Russian aggression. These numbers are still increasing daily.  

\subsection{Science for peace?}

On the occupied Ukrainian territories, starting in 2014, Russia has created ``impostor'' entities using the captured Ukrainian scientific infrastructure and integrated them into the Russian scientific community. These new entities are often named similarly or the same as the existing Ukrainian institutions-in-exile. Their purpose is to legalize and normalize appropriation of the occupied territories of Ukraine by Russia within the international scientific community. This is done through publications in which publishers or INSS do not enforce \href{https://www.iso.org/iso-3166-country-codes.html}{ISO 3166 standards} on author or journal editorial board member affiliations, or conference locations. Only during the years 2022-2024, this led to Ukrainian territories being marked as part of the Russian Federation in the Scopus database \href{https://hal.science/hal-04669865/document}{more than 2,000 times}. Today, the situation is such that the Scopus database (owned by Elsevier) locates Ukrainian cities in Russia in \href{https://hal.science/hal-04669865/document}{65\% of publications for occupied Donetsk, in 50\% for Luhansk, in 99\% for Sevastopol, and 93\% for Simferopol}. Notably over a third of those mis-affiliations is propagated through the international publishers, heavily dominated by Springer Nature/Pleiades (Germany/USA), followed by MDPI (Switzerland), EDP Science (France), IEEE (USA) and Elsevier (Netherlands). They are further spread through scientific databases (Scopus, World of Science, arxiv, \href{https://hal.science/hal-04736559/document}{inspireHEP}, etc.). Cooperation from all publishers is needed to ensure the amendment of the incorrect entries and to prevent the addition of the new ones. 

Since the start of the full-scale invasion, the Russian government has used resources of fundamental research centers (\href{https://svit.kpi.ua/en/2024/09/20/jinr-dubna-russia-an-international-laboratory-or-a-weapons-development-hub/}{JINR}, institutes of Russian Academy of Sciences, \href{https://www.federalregister.gov/documents/2022/10/04/2022-21520/additions-of-entities-to-the-entity-list}{institutes of Kurchatov NPC}, and universities) for the needs of its war against Ukraine, e.g. drone R\&D. This was discussed in \href{http://kremlin.ru/events/president/news/74277}{the June 2024 Russian government meeting} at 
\href{https://hal.science/hal-04750293/document}{JINR} chaired by Putin himself. Such programs highlight once again the need to implement measures mentioned in  \href{https://nrfu.org.ua/en/news-en/an-open-letter-from-scientists-of-ukraine-and-diaspora/}{an open letter} of the Ukrainian scientific community. 

\subsection{Actions to support the Ukrainian scientific community}
Despite the war, Ukrainian scientists have continued their research, demonstrating a strong commitment to advancing knowledge and contributing to the global scientific progress. In 2016 Ukraine joined the European Organization for Nuclear Research (CERN), as \href{https://home.web.cern.ch/news/news/cern/ukraine-becomes-associate-member-cern}{an associate member}. 
Even after the Russian full-scale invasion of 2022, the number of publications by Ukrainian scientists did not drop significantly \href{https://theconversation.com/ukrainian-science-is-struggling-threatening-long-term-economic-recovery-history-shows-ways-to-support-the-ukrainian-scientific-system-207477}{due to the support of Ukrainian scientists by the international community}.
\underline{\textbf{We list below some of the ways to support Ukrainian scientists which are particularly effective}}.

\underline{\textbf{Material support for scientists in Ukraine}} helps to minimize the brain drain and loss of infrastructure. The most impactful are programs which \textbf{foster long-term collaborations with Ukrainian institutes}. For example, \href{https://horizon-europe.org.ua/en/heo-in-ua/}{Horizon Europe office in Kyiv} was set up in 2023 to help Ukrainian scientists better integrate into international community.

One successful example of such collaboration is \href{https://ideate.lal.in2p3.fr/en/home/}{the LIA/IRP IDEATE France-Ukraine laboratory}, which fostered the Ukrainian contribution on PLUME/LHCb leading to \href{https://lhcb.web.cern.ch/Collaboration_prizes/Early_career_awards.html}{Early Career LHCb awards for 2022} for V. Zhovkovska (Orsay) and V. Orlov (CERN), originally from Ukraine. 

Another example is \href{https://www.eurizon-project.eu/news/updates/public_evaluation_report/}{the EURIZON 2023 program} of \textbf{grants for Ukrainian researchers} to perform projects in Ukraine in collaboration with the EU institutes. The program was overwhelmed to receive nearly eight-times more proposals than expected and decided to triple its budget allocation for the program. 

The CERN Council \textbf{waived the Ukrainian financial contribution} starting from the second half of 2022 and till 2024. Many international publishers allowed temporary free access to their publications and databases for scientists from Ukrainian institutes. The CERN ILO office is \textbf{\href{https://www.swissilo.ch/ilo-ukraine-initiative}{developing a program of used equipment transfer to Ukrainian institutes}}. 

\href{https://www.nature.com/articles/d41586-023-00518-y}{Since most of the Ukrainian male researchers do not have a possibility to leave Ukraine during the ongoing war}, it raises the importance of \underline{\textbf{grants for researchers based in Ukraine for remote}} \underline{\textbf{collaboration with non-Ukrainian institutes}}. There are new expanded remote initiatives \href{https://theconversation.com/ukrainian-science-is-struggling-threatening-long-term-economic-recovery-history-shows-ways-to-support-the-ukrainian-scientific-system-207477}{for Ukrainian
researchers}, for Ukrainian students (\href{https://iris-hep.org/fellows.html}{IRIS-HEP} and \href{https://jobs.smartrecruiters.com/CERN/743999952627854-ukrainian-remote-student-program}{CERN remote project programs}, CERN summer school, DESY Summer/Winter School) and teachers (CERN Ukrainian Teacher Programmes, webinars of LIA-IDEATE and Junior Academy of Sciences of Ukraine). It has recently been noted that these remote projects have a large fraction of applications from students outside the current Ukrainian particle physics centers, which indicates a potential of expanding intra-Ukrainian collaborations through such projects.

Another type of required support is 
\underline{\textbf{resident research opportunities for Ukrainian scientists}} 
\underline{\textbf{currently abroad}}. \href{https://theconversation.com/ukrainian-science-is-struggling-threatening-long-term-economic-recovery-history-shows-ways-to-support-the-ukrainian-scientific-system-207477}{``Scientists who continue to do research abroad are able to create important connections and learn new research methods that can help Ukraine transition into a more modern and internationally integrated producer of science once they return home. And if these researchers can keep professional ties with Ukraine, they may be more likely to return.''} 

Non-material assistance is most welcome too. National Research Foundation of Ukraine is looking for \href{https://nrfu.org.ua/en/contests/we-invite-experts-to-cooperate/}{experts for their grant reviews}. Peer-to-peer mentorship as well as mentorship programs are yet another example.

\section{Summary and Outlook}

Through the past century upheavals, scientists from Ukraine performed world-class research and created an extended network of scientific institutes. Our scientific community with help of international partners is working today to minimize the impact of the ongoing Russian aggression. This has led to an increased international integration of the Ukrainian scientific community.

The \href{https://indico.cern.ch/event/1395415/}{CERN-Ukraine 2024} meeting was held in Kyiv in May to jump start the preparations of the Ukrainian contribution to the update of the European Particle Physics Strategy. Discussions are also ongoing to join the FCC project at CERN. 

Louis Pasteur continued his often-quoted expression ``Science knows no country'' with ``Science is the highest personification of the nation because that nation will remain the first which carries the furthest the works of thought and intelligence.'' We are thankful to everyone who contributes to scientific life in Ukraine.  



\end{document}